\documentclass{PoS}

\makeatletter
\@addtoreset{equation}{section}

\makeatother
\usepackage{graphicx}
\usepackage{amssymb,amsfonts,amsmath}

\def\beq{\begin{equation}}
\def\eeq{\end{equation}}
\def\bea{\begin{eqnarray}}
\def\eea{\end{eqnarray}}

\newcommand{\vph}{\varphi}

\newcommand{\non}{\nonumber \\}

\title{Species Doublers as Super Multiplet Partners
in Lattice Supersymmetry }

\ShortTitle{Species Doublers as Super Multiplet Partners
in Lattice Supersymmetry}

\author{\speaker{Alessandro~D'Adda}\\
        INFN Sezione di Torino and
Dipartimento di Fisica Teorica,
Universit\`a di Torino\\
I-10125 Torino, Italy\\
        E-mail: \email{dadda@to.infn.it}}

\author{Alessandra ~Feo\\
       INFN Sezione di Torino and
Dipartimento di Fisica Teorica,
Universit\`a di Torino\\
I-10125 Torino, Italy\\
        E-mail: \email{feo@to.infn.it}}

\author{Issaku Kanamori\thanks{Current affiliation:
Institut f\"ur Theoretische Physik, Universit\"at Regensburg,
D-93040 Regensburg, Germany}
\\INFN Sezione di Torino and
Dipartimento di Fisica Teorica,
Universit\`a di Torino\\
I-10125 Torino, Italy\\
        E-mail: \email{issaku.kanamori@physik.uni-regensburg.de}}
\author{\speaker{ Noboru~Kawamoto}\\
        Department of Physics, Hokkaido University\\
Sapporo, 060-0810 Japan\\
        E-mail: \email{kawamoto@particle.sci.hokudai.ac.jp}}
 \author{Jun ~Saito\\Department of Physics, Hokkaido University\\
Sapporo, 060-0810 Japan
       \\E-mail: \email{saito@particle.sci.hokudai.ac.jp}}

\abstract{We propose a new lattice superfield formalism in momentum
representation which accommodates species doublers of the lattice
fermions and their bosonic counterparts as super multiplets. We
explicitly show that one dimensional $N=2$ model with interactions
has exact supersymmetry on the lattice for all super charges
with lattice momentum.
In coordinate representation the finite difference operator is
made to satisfy Leibnitz rule by introducing a non local  product,
the ``star'' product,  and the exact lattice supersymmetry is
realized. Supersymmetric Ward identities are shown to be satisfied
at one loop level. }

\FullConference{The XXVIII International Symposium on Lattice Field Theory, Lattice2010\\
        June 14-19, 2010\\
        Villasimius, Italy}

\begin{document}

\section{Introduction}

 There is a long history of attempts to realize exact supersymmetry
on a lattice. See \cite{latSUSYrev} for earlier and recent references.
However exact lattice supersymmetry with interactions for full extended 
supersymmetry has never been realized for gauge fields except for the 
nilpotent super charge\cite{Kaplan:2003, catterall, sugino}. It has been 
pointed out that these formulations
of lattice SUSY can be essentially reformulated by orbifolding
procedure\cite{Kaplan:2003,D-M}.

  On the other hand the link approach of lattice SUSY formulation\cite{DKKN}
includes the orbifold construction as a specific parameter choice:
shift parameter $a=0$. It was, however, claimed by several
authors \cite{Dutch, B-C-K, D-M} that an exact SUSY invariance and the
gauge invariance
are lost for non-vanishing shift parameter case of link approach:
$a\neq 0$. Then later it was recognized for non-gauge case that the
claim of the exact supersymmetry for link approach is based on the
Hopf algebraic symmetry with mild noncommutativity\cite{DKS}.

In finding a possible solution for the difficulties of the link approach,
we have found an exact lattice SUSY formulation which includes lattice
SUSY algebra in the momentum space. To show the basic ideas and explicit
presentation we examine the simplest one-dimensional $N=2$ supersymmetry
model on the lattice. The details of the formulation has already been
given in \cite{D'Adda:2010pg}. Here we explain the basic structure of the
formulation.

In the coordinate representation of the formulation we introduce a new
type of product on which the difference operator surprisingly satisfies
Leibniz rule. This new product introduces mild non-locality and 
thus compatible with a claim of no-go of lattice Leibniz rule for
difference operator in \cite{Kato:2008sp}, where another example of
exact lattice SUSY in one dimensional model is given with infinite
flavors.

\section{Basic ideas}
\label{basic-ideas}

In order to understand the basic structure of lattice SUSY, we first
consider the simplest one dimensional model with $N=1$ symmetry
in continuum theory. It is described in terms of a superfield:
\beq \Phi(x,\theta)= \varphi(x) + i \theta \psi(x), \label{superfield-0}\eeq
with a supersymmetry charge given by:
 \beq Q = \frac{\partial}{\partial \theta} +
i \theta \frac{\partial}{\partial x},~~~~~~~~~~~~~~  Q^2 = i
\frac{\partial}{\partial x}.\label{susycharge} \eeq This SUSY
algebra can be conveniently represented by introducing matrix
structure as an internal degree of freedom for super coordinate
and derivative: \beq \theta=
\begin{pmatrix}
0 & 1  \\
0 & 0
\end{pmatrix}, ~~~~~~~~~~~~~~~~~
\frac{\partial}{\partial\theta}=
\begin{pmatrix}
0 & 0  \\
1 & 0
\end{pmatrix},
\eeq
which satisfy the following anticommutation relation:
\beq
\{\frac{\partial}{\partial\theta}~,~\theta\}~=~1.
\eeq
Since this representation is not hermitian, hermiticity should be taken
care separately.

We may consider this matrix structure as an internal structure of the
space time coordinate. With respect to this internal structure the boson
$\varphi$ is considered as a field which commutes with $\theta$ and
$\frac{\partial}{\partial\theta}$ and the fermion $\psi$ as a field which
anticommutes with  them. The component fields of
boson and fermion with respect to this internal structure has then the
following form\cite{Arianos:2008ai}:
\beq
\varphi (x)=
\begin{pmatrix}
\varphi (x) & 0  \\
0 & \varphi (x)
\end{pmatrix},~~~~~~~~~~~~
 \psi (x)=
\begin{pmatrix}
\psi (x) & 0  \\
0 & -\psi (x)
\end{pmatrix}.
\label{fermion-matrix} \eeq
A super parameter may have the same internal structure as the fermion
field.

We now consider to formulate this model on the lattice. In the
matrix formulation of fields the coordinate dependence on the
lattice can be introduced by diagonal entries of a big matrix as
direct product to the internal matrix
structure~\cite{Arianos:2008ai}. It is thus very natural to
introduce half lattice structure to accommodate the $2\times 2$
matrix internal structure. One can then write a lattice
``superfield'' corresponding to (\ref{superfield-0}) as \beq
``\Phi(x)" = \varphi(x) + \frac{\sqrt{a}}{2} (-1)^\frac{2x}{a}
\psi(x) \label{lattsup}, \eeq where we have introduced a factor
$\frac {\sqrt{a}}{2}$ for later convenience and taken away the
factor $i$ for hermiticity since the second term is not a product
of two Grassmann numbers but only $\psi(x)$ is Grassmann field. In
order to accommodate hermiticity in the lattice version of SUSY
algebra (\ref{susycharge}) we need to introduce symmetric
difference operator to replace the differential operator. With
this reason we further need to introduce a quarter lattice and
then the superfield $``\Phi(x)"$ on the lattice is actually meant
as:
 \beq \Phi(x) = \left\{ \begin{array}{lc} & \varphi(x)~~~~~~~~\textrm{for}~~~
 x=n a/2 ,\\&\frac{1}{2} a^{1/2} (-1)^\frac{2x}{a}
 \psi(x) ~~~\textrm{for}~~~x=(2n+1)a/4 .
 \end{array} \right. \label{spf} \eeq
We now propose lattice supersymmetry transformations as a finite difference
over a half lattice spacing  $\frac{a}{2}$:
 \beq
 \delta\Phi(x) = \alpha a^{-1/2} (-1)^\frac{2x}{a} \left[
 \Phi(x+a/4) - \Phi(x-a/4) \right]. \label{st}
 \eeq
 By separating $\Phi(x)$ into its component fields according to
 (\ref{spf}) we find:
 \bea
&&\delta \varphi(x) = \frac{i \alpha}{2} \bigg[
\psi(x+\frac{a}{4}) + \psi(x-\frac{a}{4}) \bigg] \xrightarrow[a\to 0]{}
i \alpha \psi(x) \, ,\label{suslattf1} \\
&& \delta \psi(x) = 2 a^{-1} \alpha \bigg[ \varphi(x+\frac{a}{4})
- \varphi(x-\frac{a}{4})  \bigg]  \xrightarrow[a\to 0]{}
\alpha
\frac{\partial \varphi(x)}{\partial x} \, , \label{suslattf2} \eea
where $x$ is an even multiple of $a/4$ in (\ref{suslattf1}) and an
odd one in (\ref{suslattf2})\cite{D'Adda:2009es}.
It is surprising that the half lattice translation together with alternating
sign structure (staggered phase) for the lattice superfields generates a
correct lattice supersymmetry transformation. We consider that this
observation is a key of our formulation.

If we now introduce $N=1$ super charge as
$
\delta =\alpha Q,
$
we can show that
\beq
Q^2\varphi(x) = \frac{i}{a}\left[ \varphi(x+a/2)-\varphi(x-a/2) \right], ~~~\\
Q^2\psi(x) = \frac{i}{a}\left[ \psi(x+a/2)-\psi(x-a/2) \right].
\eeq This shows that SUSY algebra (\ref{susycharge}) is realized
in the lattice level. SUSY transformation on the lattice is half
lattice shift with alternating sign structure while super charge
square generates single lattice translation.

As we can see in the matrix representation of the component fields
$\varphi(x)$  and $\psi(x)$ in (\ref{fermion-matrix}), the same
component fields are assigned on the two neighboring half lattice
sites. It is natural to double the degrees of freedom for
component fields since we have introduced half lattice structure.
In this way it is natural to consider $N=2$ lattice SUSY
formulation to accommodate this doubled degrees of freedom. In the
following we will show that we can consistently formulate the
$D=1$, $N=2$ supersymmetry model which contains two bosonic fields
and two fermionic fields which are nicely accommodated by species
doubler states. It should also be noted that the action for
$D=N=1$ model is fermionic and thus the vacuum is not well
defined.

\section{$N=2$ supersymmetry and its lattice realization}
\label{section4}

The $N=2$, $D=1$ supersymmetry algebra has two fermionic
generators $Q_i$ satisfying the (anti)commutation relations: \beq
Q_1^2=Q_2^2= P_t, ~~~~~~\{Q_1,Q_2\}=0, ~~~~~~~~~
[P_t,Q_1]=[P_t,Q_2]=0, \label{cont-susy-alg2} \eeq where $P_t$ is
the generator of translations in the one-dimensional space-time
coordinate $t$. The superfield formulation of
(\ref{cont-susy-alg2}) makes use of two Grassmann odd fermionic
coordinates $\theta_i$ ($i=1,2$), so that the degrees of freedom
are given by the superfield expansion: \beq
\Phi(t,\theta_1,\theta_2) = \vph(t) + i \theta_1 \psi_1(t) + i
\theta_2 \psi_2(t) + i \theta_2 \theta_1 D(t), \label{superfield}
\eeq where $\psi_1$ and $\psi_2$ are Majorana fermions. The
supersymmetry transformations in terms of the component fields are
given by: \beq \delta_j \vph = i \eta_j \psi_j, ~~~~~~~ \delta_j
\psi_k=  \delta_{j,k} \eta_j \partial_t \vph + \epsilon_{jk}
\eta_j D,~~~~~~~ \delta_j D = i \epsilon_{jk} \eta_j \partial_t
\psi_k. \label{susycomp} \eeq The supersymmetric action can then
be defined in terms of the superfield $\Phi$, for instance with a
mass term and a quartic interaction: \beq \int dt d\theta_1
d\theta_2 \left[ \frac{1}{2} D_2 \Phi D_1 \Phi + i\frac{1}{2} m
\Phi^2 + i \frac{1}{4} g \Phi^4  \right]. \label{action} \eeq
 By integrating over $\theta_1$ and $\theta_2$ 
one
can obtain the action written in terms of the component fields:
\begin{align}
S=& \int dt \{ \frac{1}{2} \left[  -(\partial_t \vph)^2 - D^2 +i
\psi_1 \partial_t \psi_1 +i \psi_2 \partial_t \psi_2 \right] \non
-& m(i \psi_1 \psi_2 + D \vph) - g ( 3 i \vph^2 \psi_1 \psi_2 + D
\vph^3 ) \}. \label{actioncomp}
\end{align}

According to the discussion of the previous section a lattice of spacing $\frac{a}{2}$ is needed to represent the $N=1$ supersymmetry on the lattice.
In this way however the number of degrees of freedom are doubled. If translations are identified with shifts of $a$, a field on the lattice admits two translationally invariant configuration: the constant configuration and the configuration constant in absolute value but with alternating sign. Fluctuations around each of these configurations of a lattice field will represent two independent degrees of freedom in the continuum.
So the field content of the $N=1$ theory on the lattice, namely  one bosonic and one fermionic field, can accommodate the full field content of the $N=2$ theory. We are going to show that the $N=2$ supersymmetry algebra can be represented on the lattice in terms of  bosonic field by $\Phi(x)$ with $x=\frac{n a}{2}$, and a fermionic field $\Psi(x)$ with $x=\frac{n a}{2} + \frac{a}{4}$. As in the $N=1$ case the shift of $\frac{a}{4}$ in the fermionic field with respect
  to the bosonic one has been introduced to have symmetric finite differences in the supersymmetry transformations and
  implement hermiticity in a natural way.

 One of the two supersymmetry transformations, which we shall denote by $\delta_1$, of the $N=2$ is the same as the one already given on the lattice in the context of the $N=1$ model: \bea &&\delta_1 \Phi(x) = \frac{i \alpha}{2}
\bigg[
\Psi(x+\frac{a}{4}) + \Psi(x-\frac{a}{4}) \bigg]~~~~~~~~x=\frac{n a}{2} \, ,\label{delta1b} \\
&& \delta_1 \Psi(x) = 2  \alpha \bigg[ \Phi(x+\frac{a}{4}) -
\Phi(x-\frac{a}{4})  \bigg] ~~~~~~~~~x=\frac{n a}{2} + \frac{a}{4}
\,  \label{delta1f} \eea where $\Phi(x)$ and $\Psi(x)$ are
dimensionless, so  a rescaling of the fields with powers of $a$
will be needed to make contact with the fields of the continuum
theory. In momentum representation the supersymmetry
transformations (\ref{delta1b}) and (\ref{delta1f}) read: \beq
\delta_1 \Phi(p) =  i \cos \frac{a p}{4} \alpha \Psi(p) \, ,
~~~~~~~~~~~~~~~ \delta_1 \Psi(p) = -4 i \sin \frac{a p}{4} \alpha
\Phi(p) \, . \label{delta1pf} \eeq where for simplicity we used
the same letters for fields in momentum representation. $\Phi(p)$
and $\Psi(p)$ satisfy the following periodicity conditions: \beq
\Phi(p+\frac{4 \pi}{a} ) = \Phi(p),
\,~~~~~~~~~~~~~~~~~~\Psi(p+\frac{4 \pi}{a} ) = - \Psi(p).
\label{period} \eeq  
The commutator of two supersymmetry
transformations $\delta_1$ with parameters $\alpha$ and $\beta$
defines an infinitesimal translation $\delta^{t}_{\alpha\beta}$ of
parameter $\alpha\beta$ on the lattice: \beq
\delta^{t}_{\alpha\beta}F(p) = \delta_{1 \beta}\delta_{1
\alpha}F(p) - \delta_{1 \alpha}\delta_{1 \beta}F(p) = 4
\sin\frac{a p}{2} \alpha \beta F(p),  \label{inftr} \eeq where
$F(p)$ stands for either $\Phi(p)$ or $\Psi(p)$.
 Invariance under (\ref{inftr})  leads to a non local conservation law
where $p$ is replaced by $\sin\frac{a p}{2}$, namely , for a
product of fields of momenta $p_1$,$p_2$,...,$p_n$: \beq
\sin\frac{a p_1}{2}+\sin\frac{a p_2}{2}+ \cdots+\sin\frac{a
p_n}{2}=0 .\label{sincons} \eeq This conservation law on the
lattice was first pointed out by Dondi and Nicolai~\cite{D-N}.
In the continuum limit ($a p_i \ll1$)
(\ref{sincons}) reduces to the standard momentum conservation law
and locality is restored. The conservation law (\ref{sincons}) is
not affected if any momentum $p_i$ in it is replaced by $\frac{2
\pi}{a} - p_i$ due to the invariance of the sine. The interpretation is clear:
 in the continuum limit ( $ a p \ll 1$) $F(p) $ and $F(\frac{2 \pi}{a} - p)$
represent fluctuations of momentum $p$ respectively around the
vacuum of  momentum zero and $\frac{2 \pi}{a}$ on the lattice. So
the symmetry $p \rightarrow \frac{2 \pi}{a} - p$ amounts to
exchanging the two vacua keeping the physical momentum unchanged.

 The correspondence of the lattice fields $\Phi(p)$ and $\Psi(p)$
 with the physical fields is
\bea
&& \Phi(p) = a^{-\frac{3}{2}} \varphi(p),~~~~~~~~~~~~~~~~~~~
\Phi(\frac{2 \pi}{a} - p) = -\frac{a^{-\frac{1}{2}}}{4} D(p), \label{correspondencei}   \\
&& \Psi(p) = a^{-1} \psi_1(p),~~~~~~~~~~~~~~~~~~ \Psi(\frac{2 \pi}{a} - p) = i a^{-1} \psi_2(p) ,\label{corr1} \eea
where $p$ is  restricted in
(\ref{correspondencei},\ref{corr1}) to the interval
$(-\frac{\pi}{a}, \frac{\pi}{a})$, corresponding to a
lattice of spacing $a$.  The rescaling with powers of
$a$ keeps track of field dimensionality. With  a similar rescaling  for the
supersymmetry parameter $ \alpha = a^{-\frac{1}{2}} \eta$  we obtain for the component fields the following
transformations: \bea
 \delta_1 \varphi(p) = i \cos\frac{ap}{4} \eta \psi_1(p), &~~~~~~~~~~~~~~~~~  \delta_1 D(p) = \frac{4}{a} \sin\frac{ap}{4} \eta \psi_2(p), \label{susy1comp1}\\
 \delta_1 \psi_1(p) = -i \frac{4}{a} \sin\frac{ap}{4} \eta \varphi(p), &~~~~~~~~~~~~~~  \delta_1 \psi_2(p) = \cos\frac{ap}{4} \eta D(p). \label{susy1comp2}
 \eea
 which have the correct continuum limit.
 The component fields in (\ref{susy1comp1},\ref{susy1comp2}) are defined in the interval $-\frac{\pi}{a}<p< \frac{\pi}{a}$, hence in coordinate representation on a lattice with spacing $a$. However sine and cosine functions in (\ref{susy1comp1},\ref{susy1comp2}) are not periodic of period $\frac{2\pi}{a}$, so that SUSY transformations do not admit a local representation on such lattice.
 A lattice with $\frac{a}{2}$ spacing is needed for locality.

We have now to identify the second supersymmetry transformation $\delta_2$. In the continuum $\delta_2$ is obtained from $\delta_1$ by replacing everywhere $\psi_1(p)$ with $\psi_2(p)$, and $\psi_2(p)$ with $-\psi_1(p)$. On the lattice this corresponds to:
  \beq
  \Psi(p) \longrightarrow -i \Psi(\frac{2\pi}{a}-p).  \label{replacement}
  \eeq
  By performing this replacement on the supersymmetry transformations (\ref{delta1pf})
  one obtains the expression for $\delta_2$:
  \beq
\delta_2 \Phi(p) =   \cos \frac{a p}{4} \alpha \Psi(\frac{2\pi}{a}-p) ,
 ~~~~~~~~~~~~~~~~\delta_2 \Psi(\frac{2\pi}{a}-p) = 4  \sin \frac{a p}{4} \alpha  \Phi(p) \, . \label{delta2pf} \eeq
The supersymmetry transformation $\delta_2$   satisfies together with $\delta_1$ an $N=2$ supersymmetry algebra.  In fact the commutator of two $\delta_2$ transformations gives an infinitesimal translation as (\ref{inftr}) and the commutator of a $\delta_1$ and $\delta_2$ transformation vanishes.
  In terms of the component fields the explicit expression for the $\delta_2$ transformation can be obtained from (\ref{delta2pf}):
  \bea
 \delta_2 \varphi(p) = i \cos\frac{ap}{4} \eta \psi_2(p) & ,~~~~~~~~~~~~~ \delta_2 D(p) = -\frac{4}{a} \sin\frac{ap}{4} \eta \psi_1(p) , \label{susy2comp1}\\
 \delta_2 \psi_2(p) = -i \frac{4}{a} \sin\frac{ap}{4} \eta \varphi(p), & ~~~~~~~~~~~~~
 \delta_2 \psi_1(p) = -\cos\frac{ap}{4} \eta D(p).  \label{susy2comp2}
 \eea
 As for $\delta_1$ in the  limit $ap \ll 1$ the above transformation coincides, in the momentum space representation, with the one generated by $Q_2$ in the continuum theory.

 The coordinate representation of $\delta_2$ can be obtained directly from (\ref{susy2comp1}-\ref{susy2comp2}) by Fourier transform, or from (\ref{delta1b}-\ref{delta1f}) by performing the following substitution:
\beq \Psi(x) \longrightarrow (-1)^n
\Psi(-x)~~~~~~~~~~~~x=\frac{na}{2}-\frac{a}{4},~~~~~~~~~~~~~
\label{replacementx} \eeq which is the same as (\ref{replacement})
in the coordinate representation. Either way the result is: \bea
&&\delta_2 \Phi(x) = \frac{i \alpha}{2} (-1)^n\bigg[
\Psi(-x+\frac{a}{4}) - \Psi(-x-\frac{a}{4}) \bigg]~~~~~~~~x=\frac{n a}{2} \, ,\label{delta2b} \\
&& \delta_2 \Psi(x) = 2  \alpha (-1)^n\bigg[ \Phi(-x+\frac{a}{4})
- \Phi(-x-\frac{a}{4})  \bigg] ~~~~~~~~~x=\frac{n a}{2} +
\frac{a}{4} \, . \label{delta2f} \eea

It is clear from (\ref{delta2b}) and (\ref{delta2f}) that the
supersymmetry transformation $\delta_2$ is local in the coordinate
representation only modulo the  reflection $x \rightarrow -x$.
This was already implicit in the correspondence
(\ref{correspondencei},\ref{corr1}) between the lattice
fields and the ones of the continuum theory. In fact it is clear
from (\ref{correspondencei},\ref{corr1}) that while for
instance $\varphi(x)$ is associated to the fluctuations of
$\Phi(x)$ around the constant configuration ($p=0$), the
fluctuations of $\Phi(x)$ around the constant configuration with
alternating sign ($p=\frac{2\pi}{a}$) correspond in the continuum
to $D(-x)$. For fermions this parity change leads to a physical
meaning. Since $\psi_2(p) \leftrightarrow \psi_2(x)$ is defined as
a species doubler of $\psi_1(p) \leftrightarrow \psi_1(x)$, the
chirality of $\psi_2$ is the opposite of $\psi_1$. However by the
change of $p \rightarrow \frac{2\pi}{a}-p $ equivalently $x
\rightarrow -x$, chirality of $\psi_1$ and $\psi_2$ are adjusted
to be the same. Thus this bi local nature in the coordinate space
may be transferred to a local interpretation.

\section{Supersymmetric invariant action}

Let us define $s_1$ and $s_2$ as the supersymmetry transformations
on the lattice without the supersymmetry parameter, namely
$\delta_{i,\alpha}= \alpha s_i$. A supersymmetric invariant action
on the lattice can be defined in momentum space by giving its
$n$-point term ($n \geq 2$) in the following way:
\begin{align}
 S^{(n)}
 &=g_0^{(n)}a^n \frac{4}{n!}\int_{-\frac{\pi}{a}}^{\frac{3\pi}{a}} \frac{d p_1}{2\pi} \cdots \frac{d p_n}{2\pi}
 ~\prod_{i=1}^n\cos\frac{a p_i}{2}
   2\pi\delta\left(\sum_{i=1}^n\sin\frac{a p_i}{2}\right)
\times
  s_1s_2\Bigl[ \Phi(p_1)\Phi(p_2)\cdots\Phi(p_n)\Bigr]  \label{nterm}.
\end{align}
The sine conservation law enforced by the $\delta$ function
ensures the invariance of (\ref{nterm}) under infinitesimal
translations (\ref{inftr}) and the invariance under supersymmetry
transformations follow from the supersymmetry algebra. An explicit
evaluation of (\ref{nterm}) gives:
\begin{align}
 S_n
 &=g_0^{(n)}a^n \int_{-\frac{\pi}{a}}^{\frac{3\pi}{a}} \frac{d p_1}{2\pi} \cdots \frac{d p_n}{2\pi}
   2\pi\delta\left(\sum_{i=1}^n\sin\frac{a p_i}{2}\right)\times \prod_{i=1}^n~\cos\frac{a p_i}{2}
   \nonumber\\
 &\times \Bigl[
  2 \sin^2\frac{a p_1}{4}~\Phi(p_1)\cdots\Phi(p_n) 
  + \frac{n-1}{4}~ \sin\frac{a(p_1-p_2)}{4}~ \Psi(p_1)\Psi(p_2)\Phi(p_3)\cdots\Phi(p_n)\Bigr]. 
  \label{kterm}
\end{align}

For $n >2$ (\ref{kterm}) gives a generic interaction term, for
$n=2$ it may reproduce both kinetic and mass term. The case $n=2$
is in fact special because only in that case the sine conservation
law splits into the two separate conservation laws: \beq
p_1+p_2=0,~~~~~~~~p_1-p_2=\frac{2\pi}{a}~~~~~~~~\left( \text{mod.}\ 
\frac{4\pi}{a} \right) \label{mcon} \eeq which are linear in the
momenta. The delta function in (\ref{kterm}) can then be replaced
without loss of invariance by a superposition with arbitrary
coefficients delta functions, namely: \beq a
g_0^{(2)}\prod_{i=1}^2\cos\frac{ap_i}{2} \delta(\sin\frac{a
p_1}{2}+\sin\frac{a p_2}{2}) \longrightarrow \delta(p_1+p_2) + m_0
\delta(p_1-p_2-\frac{2\pi}{a}) \label{deltas} \eeq where $m_0$ is
a free parameter. The first delta function in (\ref{deltas})
generates  the supersymmetric kinetic term, the second delta
function the supersymmetric mass term given respectively by: \bea
S_{kin} &=& a \int_{-\frac{\pi}{a}}^{\frac{3\pi}{a}} \frac{dp}{2\pi}
\left[ 2 \sin^2\frac{ap}{4} \Phi(-p)\Phi(p) - \frac{1}{4} \sin
\frac{ap}{2} \Psi(-p)\Psi(p)\right],           \label{skin}\\ 
S_{mass} &=& a m_0 \int_{-\frac{\pi}{a}}^{\frac{3\pi}{a}}
\frac{dp}{2\pi} \left[ \Phi(p+\frac{2\pi}{a})\Phi(p) + \frac{1}{4}
\Psi(p+\frac{2\pi}{a})\Psi(p)\right]. \label{smass} \eea
We can now
use the correspondence (\ref{correspondencei},\ref{corr1}) to
write (\ref{skin}) and (\ref{smass}) in terms of the component
fields, defined on a $\frac{2\pi}{a}$ interval: 
\bea S_{kin} 
&=&\int_{-\frac{\pi}{a}}^{\frac{\pi}{a}} \frac{d p}{2\pi} \Bigl[
\frac{4}{a^2} (1-\cos\frac{a p}{2}) \varphi(-p)\varphi(p)
+\frac{1}{4}(1+\cos\frac{a p}{2}) D(-p)D(p)-\nonumber \\ &&-
\frac{1}{a} \sin\frac{a p}{2}\psi_1(-p)\psi_1(p)-\frac{1}{a}
\sin\frac{a p}{2}\psi_2(-p)\psi_2(p)  \Bigr].\label{kincomp} \eea
Similarly for the mass term we get: \beq S_{mass} = 2 m
\int_{-\frac{\pi}{a}}^{\frac{\pi}{a}} \frac{d p}{2\pi} \left[
-\varphi(-p)D(p) -i \psi_1(-p)\psi_2(p) \right], \label{masscomp}
\eeq where $m$ is now the physical mass: $m = \frac{m_0}{a}$.
Thanks to the rescaling all fields in (\ref{kincomp}) and
(\ref{masscomp}) have the correct canonical dimension, and the
continuum limit is smooth. The component fields
$\varphi(p)$,$D(p)$,$\psi_1(p)$ and $\psi_2(p)$ are defined for
$p$ in the interval $(-\frac{\pi}{a},\frac{\pi}{a})$. This is the
Brillouin zone corresponding to a lattice of spacing $a$, so we
could define a lattice with coordinates $\tilde{x}= n a$ and the
component fields on it as the Fourier transforms of the momentum
space components. However the action written in the coordinate
$\tilde{x}$ space is non-local, since the finite difference
operators appearing in (\ref{kincomp}) are periodic with period
$\frac{4 \pi}{a}$ and not $\frac{2\pi}{a}$ that would be needed
for a local expression on a lattice with spacing $a$. Instead it
is possible to write (\ref{skin}) and (\ref{smass}) as  local
actions
  (modulo $x \rightarrow -x$ couplings) on the lattice of spacing $\frac{a}{2}$.

The kinetic term can be written as: \beq S_{kin} = \frac{1}{4}
\sum_{x=n\frac{a}{2}} \left[ \Phi(x) \left( 2 \Phi(x) -
\Phi(x+\frac{a}{2})-\Phi(x-\frac{a}{2}) \right) + \frac{i}{2}
\Psi(x+\frac{a}{4})\Psi(x-\frac{a}{4}) \right]. \label{coordskin2}
\eeq The coordinate representation for the mass term (\ref{smass})
shows instead a coupling between fields in $x$ and $-x$: \beq
S_{mass}= \frac{m_0}{2} \sum_{x=n\frac{a}{2}} (-1)^{\frac{2x}{a}}
\left[ \Phi(-x)\Phi(x) + \frac{i}{4}
\Psi(-x-\frac{a}{4})\Psi(x+\frac{a}{4}) \right].
\label{coordsmass} \eeq The bi local structure of
(\ref{coordsmass}) shows that the extended lattice  with spacing
$\frac{a}{2}$ has not a straightforward relation to the coordinate
space in the continuum limit. This is related to the fact that
while the fluctuations of $\Phi(x)$ (resp. $\Psi(x)$) around a
constant field configuration are associated to the component field
$\varphi(x)$ (resp. $\psi_1(x)$), its fluctuations around
$(-1)^{\frac{2 x}{a}}$ are associated to $D(-x)$ (resp.
$\psi_2(-x)$). In other words the two bosonic (resp. fermionic)
components of the superfield are embedded in a single bosonic
(resp. fermionic) field on the extended lattice  is non trivial
and exhibits a bi local structure. Although the extended lattice
is not a discrete representation of superspace (bosonic and
fermionic fields have to be introduced separately on it) it
carries some information about the superspace structure and as
such it does not simply map onto the coordinate space in the
continuum limit.

For $n > 2$ the general invariant expression (\ref{kterm})
describes interaction terms. The sine conservation law is in these
cases intrinsically non-linear in the momenta, and consequently
the interaction terms are non-local in coordinate representation.
However, as it is shown in the following section, they can be
formulated in terms of a non local (but still associative and
commutative)product, which we name ``star product'' , in place of
the ordinary field product. As for the kinetic and the mass term
the interaction terms can be expressed in terms of the component
fields by using the correspondence (\ref{correspondencei},
\ref{corr1})and splitting each momentum integration into the two
regions $(-\frac{\pi}{a}, \frac{\pi}{a})$ and $(\frac{\pi}{a},
\frac{3\pi}{a})$. Since each bosonic field $\Phi(p)$ can represent
either $\varphi$ or $D$, depending on the value of $p$, the
expansion in terms of component fields produces a large number of
terms including couplings that do not appear in the continuum
limit. Due to the different rescaling of the fields these terms
have different powers of $a$, and by carefully counting the powers
of $a$ one finds that in order to have a finite and non vanishing
continuum limit for the leading term of (\ref{nterm}) the physical
coupling constant $g^{(n)}$ must be defined as: 
\beq g^{(n)}=
a^{-\frac{n}{2}} g_0^{n}. \label{coupling} \eeq 
The naive
continuum limit of the lattice theory can then be taken, it is
smooth and it reproduces the continuum supersymmetric theory
described for instance, for a $\Phi^4$ interaction, in eq.
(\ref{actioncomp}). This however is not sufficient: invariance
under finite translations is violated on the lattice by the sine
conservation law. It is then crucial that translational invariance
is recovered in the continuum limit. This is not obvious and it
requires the analysis of the UV properties of the theory when the
continuum limit is taken. The lattice theory described in the
previous section  in terms of the fields $\Phi$ and $\Psi$ is free
of ultraviolet divergences. In fact everything in that theory can
be written in terms of the dimensionless  momentum variables
$\tilde{p}_i= a p_i$, which are
 angular variables with periodicity $ 4\pi$. Momentum
integrations reduce to integrations over trigonometric functions
of $\tilde{p}_i$, and ultraviolet divergences never appear. All
correlation functions of $\Phi$s and $\Psi$s integrations are
therefore finite. This however is not enough to ensure that the
continuum limit is smooth and that ultraviolet divergences do not
appear in the limiting process. The continuum limit in fact
involves  a rescaling of fields with powers of $a$, which is
singular in the $a \rightarrow 0$ limit. At the same time the
continuum limit, being a limit where $a \rightarrow 0$ keeping the
physical momentum fixed, corresponds to a situation where all
external momenta $\tilde{p}_i$ are in the neighborhood of one of
the vacua, namely at $\tilde{p}_i=0$ or $\tilde{p}_i=2 \pi$. The
limit being a singular one, the ultraviolet behavior has to be
checked.
This was done in ref. ~\cite{D'Adda:2010pg} where the lack of UV divergences in the continuum limit was explicitly checked.
The recovery of translational invariance in the continuum limit can then be verified, as shown below.
 Since the conserved quantity on the lattice is not the momentum itself $p$ but $\sin\frac{ap}{2}$
finite translational invariance is explicitly broken at a finite
lattice spacing. Indeed, if we denote the component fields by
$\phi_A = (\varphi, D, \psi_1, \psi_2)$, the sine conservation law
implies that correlation functions are invariant under the
transformation:
\begin{align}
 \phi_A(p) &\rightarrow \exp(i l \frac{2}{a}\sin\frac{ap}{2}) \phi_A(p)
 \qquad l:\mbox{a finite length}
 \label{eq:finite-translation-sin}
\end{align}
whereas invariance under finite translation would require the invariance under the transformation
\begin{align}
 \phi_A(p) &\rightarrow \exp(i lp)\phi_A(p).\label{fint}
\end{align}
To prove that invariance under finite translations is recovered we need to
prove that in the continuum limit (\ref{eq:finite-translation-sin}) and (\ref{fint}) are equivalent.
For an $n$-point correlation function of $\phi_A $,
transformation (\ref{eq:finite-translation-sin}) is equivalent to
\begin{align}
 &\langle \phi_{A_1}(p_1) \phi_{A_2}(p_2) \dots \phi_{A_n}(p_n) \rangle
 \rightarrow
  \exp\left(\sum_{i=1}^n \frac{2l}{a} \sin\frac{ap}{2}\right)
 \langle \phi_{A_1}(p_1) \phi_{A_2}(p_2) \dots \phi_{A_n}(p_n) \rangle \nonumber \\
 &\simeq
 \left( 1 -i \frac{a^2l}{24}\sum_{i=1}^{n}p_i^3\right)
 \exp\left(il\sum_{i=1}^n p_i\right)
 \langle \phi_{A_1}(p_1) \phi_{A_2}(p_2) \dots \phi_{A_n}(p_n) \rangle, \label{compr}
\end{align}
where in the last step higher order terms in the expansion of $\sin\frac{ap}{2}$ have been neglected since $ap \rightarrow 0$ in the continuum limit.
The leading term that breaks translational invariance is then given by the second term in the bracket at the r.h.s. of (\ref{compr}). This vanishes as $a^2$ in the continuum limit if we assume $l p_i$ to be of order $1$ so that this term can be neglected as long as no divergence ( of order at least $\frac{1}{a^2}$ ) arises in the correlation  function $\langle \phi_{A_1}(p_1) \phi_{A_2}(p_2) \dots \phi_{A_n}(p_n) \rangle$.
As discussed before this is not the case, so we can conclude that invariance under finite translations is recovered in the continuum limit.

Finally let us consider invariance under supersymmetry.
Invariance under supersymmetry transformations is exact at the
finite lattice spacing and it is not spoiled by radiative
corrections, which are all finite in the lattice theory. Since the
continuum limit is smooth, we expect that  exact supersymmetry is
preserved also in this limit. This can be confirmed explicitly, as shown in ~\cite{D'Adda:2010pg}, by checking that the corresponding Ward-Takahashi identities (WTi) are satisfied.

\section{Coordinate representation}
\label{coordinate}

As we have seen in the previous section, momentum conservation on the lattice
is the sine momentum conservation. We now try to find out the coordinate
counterpart of the corresponding formulation.

With ordinary momentum conservation the
product of a field $F$ of momentum $p_1$ and a field $G$
of momentum $p_2$ is a composite field $\Phi=F \cdot G$
of momentum $p=p_1+p_2$, namely the momentum is the additive
quantity under product: \beq \Phi(p) \equiv (F\cdot G)(p)
= \frac{a}{2\pi}\int d p_1 d p_2 F(p_1) G(p_2)
\delta(p-p_1-p_2). \label{dotprod} \eeq
 In coordinate space this amounts to the ordinary local product:
 \beq
 \Phi(x) \equiv (F \cdot G)(x) = F(x) G(x).
 \label{dotcoord}
 \eeq
 On the lattice momentum conservation is replaced by the
lattice (sine) momentum conservation (\ref{sincons}), which means that $\hat{p}= \frac{2}{a} \sin\frac{a p}{2}$ is the additive quantity when taking the product of two fields. In other words  the product of a field $F$ of momentum $p_1$ and a field $G$ of momentum $p_2$ is a composite field $\Phi=F * G$ of momentum $p$ with $\sin\frac{ap}{2}=\sin\frac{a p_1}{2}+\sin\frac{a p_2}{2}$.
 This amounts to changing the definition of the ``dot'' product  to that of a ``star'' product defined in momentum space as:
 \beq
 \Phi(p) \equiv (F * G)(p)
= \frac{a}{2\pi}\int d \hat{p}_1 d \hat{p}_2 F(p_1) G(p_2) \delta(\hat{p}-\hat{p}_1-\hat{p}_2)   \label{starprod}
\eeq As we shall see this product is not anymore local in
coordinate space but satisfies the Leibniz rule with respect to
the symmetric difference operator $\hat{\partial}$. This is easily
checked in the momentum representation. In fact  acting with the symmetric difference operator corresponds in
momentum space to multiplication by
$\hat{p}=\frac{2}{a}\sin\frac{ap}{2}$, so that from
 (\ref{starprod}) we get: \beq \hat{p}~ \Phi(p)
 = \frac{a}{2\pi}\int ~d \hat{p}_1~ d \hat{p}_2 \left[ \hat{p}_1~ F(p_1) ~
G(p_2)+ F(p_1)~\hat{p}_2~G(p_2) \right]
\delta(\hat{p}-\hat{p}_1-\hat{p}_2).   \label{lbrule} \eeq

Explicit form of the coordinate representation of the star product
is given by
\begin{align}
(F*G) (x)
&= F(x) * G(x)
=a\int \frac{d\hat{p}}{2\pi}\, e^{-ipx}~ (F*G)(p) \nonumber \\
&=\int_{-\frac{\pi}{2}}^{\frac{3\pi}{2}} d\tilde{p}~ \cos\tilde{p}
~e^{-ipx} \int_{-\frac{\pi}{2}}^{\frac{3\pi}{2}}
\frac{d\tilde{p}_1}{2\pi} \frac{d\tilde{p}_2}{2\pi}
\cos\tilde{p}_1 ~\cos\tilde{p}_2
\int_{-\infty}^\infty  \frac{d\tau}{2\pi}
e^{i{\tau}(\sin\tilde{p}-\sin\tilde{p}_1-\sin\tilde{p}_2)}
\nonumber \\
&~~~~~~~\times \sum_{y,z}e^{i(m\tilde{p}_1 +l\tilde{p}_2)} F(y) G(z)
\nonumber \\
&= \int_{-\infty}^\infty d\tau J_{n\pm1}({\tau})
\sum_{m,l}J_{m\pm1}({\tau})J_{l\pm1}(\tau) F(y) G(z),
\label{def-star-prod-cor}
\end{align}
where $\tilde{p}=\frac{ap}{2}$, 
and $x=\frac{na}{2}, y=\frac{ma}{2}, z=\frac{la}{2}$ should be
understood. The lattice delta function is parameterized by $\tau$
\begin{align}
\delta\left(\frac{2}{a}\sin\tilde{p}_i\right)
=\frac{a}{4\pi}\int^\infty_{-\infty}
d\tau e^{i{\tau} \sin\tilde{p}_i}.
\label{deltafunction}
\end{align}
$J_n(\tau)$ is a Bessel function defined as
\begin{align}
J_n(\tau)=\frac{1}{2\pi}\int_\alpha^{2\pi+\alpha}
e^{i(n\theta-\tau \sin \theta)} d\theta,
\label{Bessel}
\end{align}
and we use the following notation:
\begin{align}
J_{n\pm1}(\tau)=\frac{1}{2}(J_{n+1}(\tau) + J_{n-1}(\tau)).
\label{def-Jpm}
\end{align}
It is obvious that the star product is commutative:
\beq
F(x) * G(x) = G(x) * F(x).
\label{star-commut}
\eeq
We can now check how the difference operator acts on the star product
of two lattice superfields and find that the difference operator action
on a star product indeed satisfies Leibniz rule:
\begin{align}
\lefteqn{
 i\hat{\partial} (F(x)*G(x))
} \quad \nonumber\\ 
&=
a\int \frac{d\hat{p}}{2\pi}~ i\hat{\partial}_x~e^{-ipx}~(F*G)(p)
\nonumber \\
&=\frac{a^2}{4} \int d\hat{p}~e^{-ipx}\sum_{y,z}\int
\frac{d\hat{p}_1}{2\pi} \frac{d\hat{p}_2}{2\pi}~e^{ip_1y+ip_2z}
\left((i\hat{\partial}_y~F(y))G(z)~+~
F(y)~(i\hat{\partial}_z~G(z)) \right)
~\delta (\hat{p}-\hat{p}_1-\hat{p}_2)\nonumber \\
&=(i\hat{\partial} F(x))*G(x) + F(x) * (i\hat{\partial} G(x)).
\label{star-Leibniz}
\end{align}

One can then show that this definition of the $*$-product leads to the
vanishing nature of surface terms for $*$-products of several
fields:
\begin{align}
\lefteqn{
\sum_x i\hat{\partial}( F(x)*G(x)*H(x)*\cdots)
} \nonumber \quad \\
&=\sum_x \left( (i\hat{\partial}F(x) )*G(x)*H(x)+
F(x)*(i\hat{\partial}G(x))*H(x) \right. 
+\left. F(x)*G(x)*(i\hat{\partial}H(x) ) + \cdots \right)
= 0.
\label{surface-3prod}
\end{align}
It should be noted that if the $*$-product in the above definition is simply
replaced by the normal product, the surface terms does
not vanish for products of fields more than three fields.

We can now generalize the definition of $*$-product for a product of
any fields of the form:
\begin{align}
F_1(x+b_1)* F_2(x+b_2) * \cdots *F_n(x+b_n) = 
 \int^\infty_{-\infty} d\tau
J_{n\pm1}(\tau)\sum_{m_1,\cdots,m_n}
\left(\prod_{j=1}^n J_{m_j\pm1}(\tau)F_j(y_j+b_j)\right),
\label{star-n-prod}
\end{align}
where $x=\frac{na}{2}$, $y_j=\frac{m_j a}{2}$ and $J_{n\pm1}(\tau)$ is
defined in (\ref{def-Jpm}).

We can now derive the coordinate representation of the
general interaction action $S^{(n)}$.
We first note the following relation:
\begin{align}
2\pi\delta\left(\sum_{j=1}^n\sin\frac{a p_j}{2}\right)
&= \frac{a}{2}\sum_x\int d(\sin\tilde{p}) e^{-ipx}\delta\left( \sin\tilde{p}
-\sum_{j=1}^n\sin\tilde{p}_j \right).
\label{delta-relation}
\end{align}
Then the general interaction action (\ref{nterm}) can be expressed by the
$*$-products:
\begin{align}
 S^{(n)}
&=\frac{4}{n!}g_0^{(n)} \sum_x
\Bigl[\left(2\Phi(x)
- \Phi(x+\frac{a}{2})
-\Phi(x-\frac{a}{2})\right)* \Phi^{n-1}(x)
\nonumber \\
&~~~~~~~~~~~~~~~~~~~~~~~~~~~~~~~+\frac{(n-1)i}{2} \Psi(x+\frac{3a}{4})*
\Psi(x+\frac{a}{4})* \Phi^{n-2}(x) \Bigr],
\label{star-Sn-action}
\end{align}
where $\Phi^{n-1}(x)$ is $(n-1)$-th power of $*$-product.
The $S^{(2)}$ action in the star products form is equivalent
to a sum of both the kinetic terms and the mass terms with fixed
coefficient, which include product of local fields and has the
following form:
\begin{align}
 S^{(2)}
&= \sum_{x}\Bigl[\Phi(x)* \left(2\Phi(x)
- \Phi(x+\frac{a}{2})-\Phi(x-\frac{a}{2})\right)
+\frac{i}{2} \Psi(x+\frac{3a}{4}) * \Psi(x+\frac{a}{4})\Bigr].
  \label{S2*term}
\end{align}
It is interesting to recognize that the coordinate representation
of the action with star product has almost the same form of the
kinetic term of the local action, $S_{kin}$ in (\ref{coordskin2}),
where the star product is just replaced by the normal product. The
arguments of the fermionic lattice superfield in $S_{kin}$ is
shifted with $\frac{a}{2}$ from that of (\ref{S2*term}). This is
due to the loss of lattice translational invariance in the star
product formulation while in eq. (\ref{coordskin2}) the
translation w.r.t $\frac{a}{2}$ shift is recovered (this is a
special feature of bi linear terms) and thus we can obtain the
same arguments.

The non local nature of the star product should disappear in the
continuum limit. This is however non trivial due to the $p
\rightarrow \frac{2 \pi}{a} - p$ symmetry of the $\sin\frac{a
p}{2}$ function and the existence of two translationally invariant
vacua at $p=0$ and $p=\frac{2\pi}{a}$. It was shown by Dondi and
Nicolai~\cite{D-N} that in the continuum limit namely at fixed $x$
with $a\rightarrow 0$: \beq J_{\frac{2 x}{a}} (\tau) \rightarrow
\delta(\tau-\frac{2 x}{a}).
  \label{contlimit}
\eeq However in the present context the continuum limit picks up
also the configuration at $p= \frac{2\pi}{a}$ and the previous
result has to be replaced by: \beq J_{\frac{2 x}{a}} (\tau)
\rightarrow \delta(\tau-\frac{2 x}{a})+ (-1)^{\frac{2
x}{a}}\delta(\tau+\frac{2 x}{a}). \label{contlim2} \eeq Thus
locality is recovered in the continuum limit, but with an extra
coupling of fields in the points $x$ and $-x$ accompanied with the
alternating sign factor $(-1)^{\frac{2x}{a}}$. Such remaining non
locality disappears when the lattice field $\Phi$ and $\Psi$ are
reinterpreted in terms of component fields.

\section{Conclusion and discussions}

We have proposed a new lattice supersymmetry formulation which ensures an
exact Lie algebraic supersymmetry invariance on the lattice for all super
charges even with interactions.
We have introduced bosonic and fermionic lattice superfields which accommodate
species doublers as bosonic and fermionic particle fields of super multiplets.

As the simplest model we have explicitly investigated $N=2$ model
in one dimension. The model includes interaction terms and the
exact lattice supersymmetry invariance of the action for two
supersymmetry charges with lattice momentum are shown explicitly.
In the momentum representation of the formulation the standard momentum
conservation is replaced by the lattice counterpart of momentum
conservation: the sine momentum conservation. The basic lattice
structure of this one dimensional model is half lattice spacing
structure and the lattice supersymmetry transformation is
essentially a half lattice spacing translation. The super
coordinate structure and the momentum representation of species
doubler fields is hidden implicitly in the alternating sign
structure of a half lattice spacing in the coordinate space.

Since the symmetric difference operator does not satisfy Leibniz rule,
it was very natural to ask how the supersymmetry algebra be consistent
in the coordinate space since super charges satisfy Leibniz rule.
In the link approach this problem was avoided by introducing shift
nature for super charges. In the current formulation this puzzle
is beautifully solved by introducing a new star product of lattice
superfields: The difference operator satisfies Leibniz rule on
the star products of lattice super fields.

In the definition of the star product in (\ref{def-star-prod-cor}),
non-local summation is introduced with Bessel functions. This
non-local feature is not well-behaved non-locality as
exponential type but not worse than the inverse distance behaviour. It is
expected to be in-between of these two behaviours since Bessel function
is integrable. Our claim in this paper includes the statement that exact
supersymmetry on the lattice accompanies a non-local behaviour.

Since we have established a new lattice supersymmetry formulation
which has exact supersymmetry on the lattice, it would be important
to extend the formulation into higher dimensions and to the models
with gauge fields. An extension to two dimensions will be given
elsewhere.



\subsection*{Acknowledgments}
This work was supported in part by Japanese Ministry of Education,
Science, Sports and Culture under the grant number 50169778 and
also by Insituto Nazionale di Fisica Nucleare (INFN) research
funds. I.K. was financially supported by Nishina memorial
foundation.

\end{document}